\documentclass[sigconf]{acmart}
\usepackage{multirow}
\usepackage{enumitem}
\setitemize{leftmargin=*}

\AtBeginDocument{%
  \providecommand\BibTeX{{%
    \normalfont B\kern-0.5em{\scshape i\kern-0.25em b}\kern-0.8em\TeX}}}

\copyrightyear{2024}
\acmYear{2024}
\setcopyright{acmlicensed}\acmConference[SIGIR '24]{Proceedings of the 47th
International ACM SIGIR Conference on Research and Development in
Information Retrieval}{July 14--18, 2024}{Washington, DC, USA}
\acmBooktitle{Proceedings of the 47th International ACM SIGIR Conference on
Research and Development in Information Retrieval (SIGIR '24), July 14--18,
2024, Washington, DC, USA}
\acmDOI{10.1145/3626772.3657724}
\acmISBN{979-8-4007-0431-4/24/07}
\begin{document}

\title[AFDGCF: Adaptive Feature De-correlation Graph Collaborative Filtering for Recommendations]{AFDGCF: Adaptive Feature De-correlation Graph Collaborative Filtering for Recommendations}

\author{Wei Wu}
\affiliation{
  \institution{School of Data Science, University of Science and Technology of China}
  \city{Hefei}
  \country{China}
}
\email{urara@mail.ustc.edu.cn}

\author{Chao Wang}
\authornote{Corresponding authors}
\affiliation{
  \institution{Guangzhou HKUST Fok Ying Tung Research Institute}
  \city{Guangzhou}
  \country{China}
  \and
  \institution{School of Computer Science and Technology, University of Science and Technology of China}
  \city{Hefei}
  \country{China}
}
\email{chadwang2012@gmail.com}

\author{Dazhong	Shen}
\affiliation{
  \institution{Shanghai Artificial Intelligence Laboratory}
  \city{Shanghai}
  \country{China}
}
\email{dazh.shen@gmail.com}

\author{Chuan Qin}
\affiliation{
  \institution{Career Science Lab, BOSS Zhipin} 
  \city{Beijing} 
  \country{China} 
  \and 
  \institution{PBC School of Finance, Tsinghua University} 
  \city{Beijing} 
  \country{China}}
\email{chuanqin0426@gmail.com}

\author{Liyi Chen}
\affiliation{
  \institution{School of Data Science, University of Science and Technology of China}
  \city{Hefei}
  \country{China}
}
\email{liyichencly@gmail.com}

\author{Hui Xiong}
\authornotemark[1]
\affiliation{
  \institution{Thrust of Artificial Intelligence, The Hong Kong University of Science and Technology (Guangzhou)}
  \city{Guangzhou}
  \country{China} \and
  \institution{Department of Computer Science and Engineering, The Hong Kong University of Science and Technology}
  \city{Hong Kong SAR}
  \country{China}
}
\email{xionghui@ust.hk}
\renewcommand{\shortauthors}{Wei Wu, et al.}


\begin{abstract}
Collaborative filtering methods based on graph neural networks (GNNs) have witnessed significant success in recommender systems (RS), capitalizing on their ability to capture collaborative signals within intricate user-item relationships via message-passing mechanisms. However, these GNN-based RS inadvertently introduce excess linear correlation between user and item embeddings, contradicting the goal of providing personalized recommendations. While existing research predominantly ascribes this flaw to the over-smoothing problem, this paper underscores the critical, often overlooked role of the over-correlation issue in diminishing the effectiveness of GNN representations and subsequent recommendation performance. Up to now, the over-correlation issue remains unexplored in RS. Meanwhile, how to mitigate the impact of over-correlation while preserving collaborative filtering signals is a significant challenge.
To this end, this paper aims to address the aforementioned gap by undertaking a comprehensive study of the over-correlation issue in graph collaborative filtering models. Firstly, we present empirical evidence to demonstrate the widespread prevalence of over-correlation in these models. Subsequently, we dive into a theoretical analysis which establishes a pivotal connection between the over-correlation and over-smoothing issues. Leveraging these insights, we introduce the \underline{\textbf{A}}daptive \underline{\textbf{F}}eature \underline{\textbf{D}}e-correlation \underline{\textbf{G}}raph \underline{\textbf{C}}ollaborative \underline{\textbf{F}}iltering (AFDGCF) framework, which dynamically applies correlation penalties to the feature dimensions of the representation matrix, effectively alleviating both over-correlation and over-smoothing issues.
The efficacy of the proposed framework is corroborated through extensive experiments conducted with four representative graph collaborative filtering models across four publicly available datasets. Our results show the superiority of AFDGCF in enhancing the performance landscape of graph collaborative filtering models.
The code is available at \url{https://github.com/U-rara/AFDGCF}.
\end{abstract}





\maketitle

\section{Introduction}
Recommender system (RS) has become widely adopted in various online applications as a solution to the challenge of information overload, by providing personalized item recommendations~\cite{wu2022survey}. Among the popular RS techniques, collaborative filtering (CF) stands out for its ability to extract informative user and item representations from historical interactions~\cite{su2009survey}. With the rapid advancement of graph neural networks (GNNs) and their effectiveness in processing graph-structured data, GNN-based CF has emerged as a prominent research focus in RS, capturing collaborative signals in the high-order connectivity between users and items~\cite{wang2019neural}. Despite the remarkable achievements facilitated by GNNs in RS, significant challenges persist in attaining high robustness and accuracy, such as feature over-correlation and over-smoothing problems.

In the context of GNN-based CF models, a single GNN layer predominantly considers the immediate neighboring nodes of users and items, potentially limiting the ability to capture deep collaborative signals. Addressing this limitation, conventional GNN-based CF models stack multiple GNN layers to expand their receptive fields. However, this practice can lead to performance degradation as the number of stacked layers increases~\cite{wang2019neural,zhao2019pairnorm}. Prior studies on GNN-based CF~\cite{zhao2019pairnorm,liu2020towards,chen2020measuring,rusch2023survey} often attribute this degradation to the widely discussed over-smoothing problem, where node representations tend to converge towards similarity with escalating layer count. However, it is important to note that another critical factor contributing to performance decline is the feature over-correlation issue. As the number of GNN layers grows, the dimensions of node representations' features become increasingly correlated~\cite{jin2022feature}, negatively impacting the quality of the learned embeddings. While over-correlation and over-smoothing exhibit similar tendencies and effects, their primary distinction lies in their focus on relationships: over-smoothing pertains to relationships between node representations (in the row direction of the representation matrix), while over-correlation pertains to relationships between feature dimensions of the representations (in the column direction of the representation matrix).

In the literature, while various GNN-based CF studies, including GCMC~\cite{berg2017graph}, SpectralCF~\cite{zheng2018spectral}, PinSage~\cite{ying2018graph}, and NGCF~\cite{ngcf}, have primarily concentrated on GNN structures, they have often overlooked the aforementioned issues. Subsequent investigations~\cite{he2020lightgcn,chen2020revisiting,mao2021ultragcn,liu2021interest,peng2022svd,xia2022hypergraph,cai2023lightgcl,xia2023graph} have started addressing these problems, with an emphasis on mitigating the over-smoothing issue but largely neglecting feature over-correlation. Consequently, several unresolved challenges remain in GNN-based CF. Firstly, the influence of feature over-correlation on RS and its relationship with over-smoothing remains unclear. If both issues concurrently impact the recommendation performance, it may be possible to leverage their interconnection to address both problems simultaneously. Secondly, devising dedicated strategies to alleviate feature over-correlation in GNN-based CF models remains an unexplored but essential avenue for enhancing embedding capability. Lastly, the varying severity of over-correlation and over-smoothing across different GNN layers underscores the importance of crafting layer-wise adaptive techniques to appropriately control feature learning.

To address these challenges, in this paper, we propose a comprehensive de-correlation paradigm for GNN-based CF models, named the \underline{\textbf{A}}daptive \underline{\textbf{F}}eature \underline{\textbf{D}}e-correlation \underline{\textbf{G}}raph \underline{\textbf{C}}ollaborative \underline{\textbf{F}}iltering (AFDGCF) framework. To the best of our knowledge, this is the first investigation into the impact of feature over-correlation issue in RS. We commence by providing empirical validation, demonstrating the prevalence of the feature over-correlation issue in GNN-based methods. Subsequently, through theoretical analysis, we establish a fundamental connection between feature over-correlation and over-smoothing. Building upon this insight, we propose the model-agnostic AFDGCF framework, which adaptively penalizes correlations among feature dimensions within the output representations of each GNN layer to mitigate the impact
of over-correlation while preserving collaborative filtering signal. The efficacy of our approach is extensively validated through the experiments on different implementations employing diverse GNN-based CF methods and various publicly available datasets. These results demonstrate the general effectiveness of our AFDGCF framework in alleviating both the over-correlation and over-smoothing issues, and thus enhancing model performance.
In summary, the main contributions of this paper are as follows: 1) We introduce a new perspective on the feature over-correlation problem in GNN-based CF. We theoretically discuss the relevance and differences between over-correlation and the commonly addressed over-smoothing problem. 2) We propose the Adaptive Feature De-correlation Graph Collaborative Filtering (AFDGCF) framework, a novel model-agnostic approach to mitigate feature over-correlation issue adaptively in existing GNN-based CF. 3) We present extensive experimental results using various public datasets to validate the effectiveness of the AFDGCF framework compared to many state-of-the-art methods.

\begin{figure}[t]
\centering\includegraphics[width=0.47\textwidth]{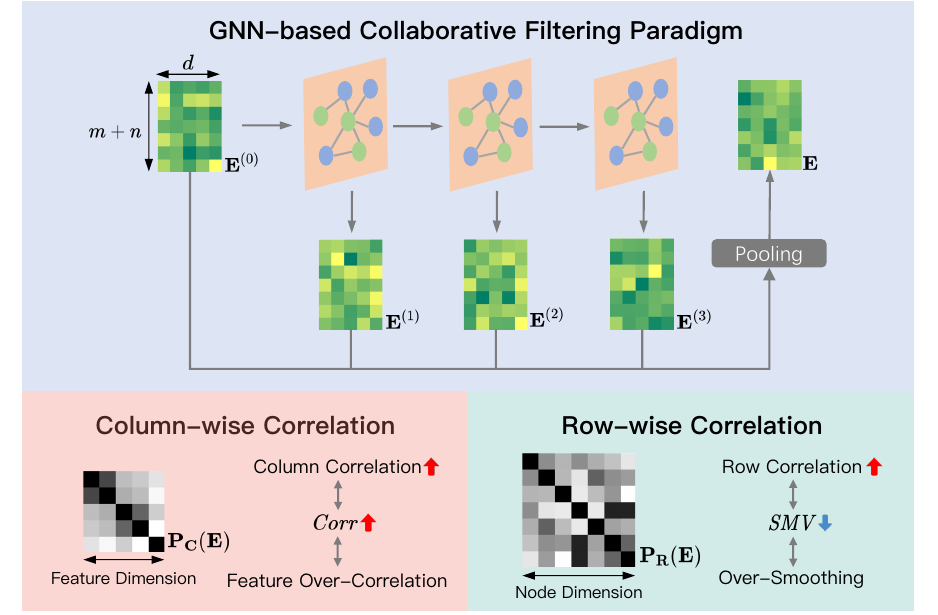}
\caption{Illustration of the GNN-based CF, highlighting the challenges of over-correlation and over-smoothing.}
\label{fig:illu}
\vspace{-5mm}
\end{figure}
\begin{figure*}[!ht]
\centering\includegraphics[width=1\textwidth]{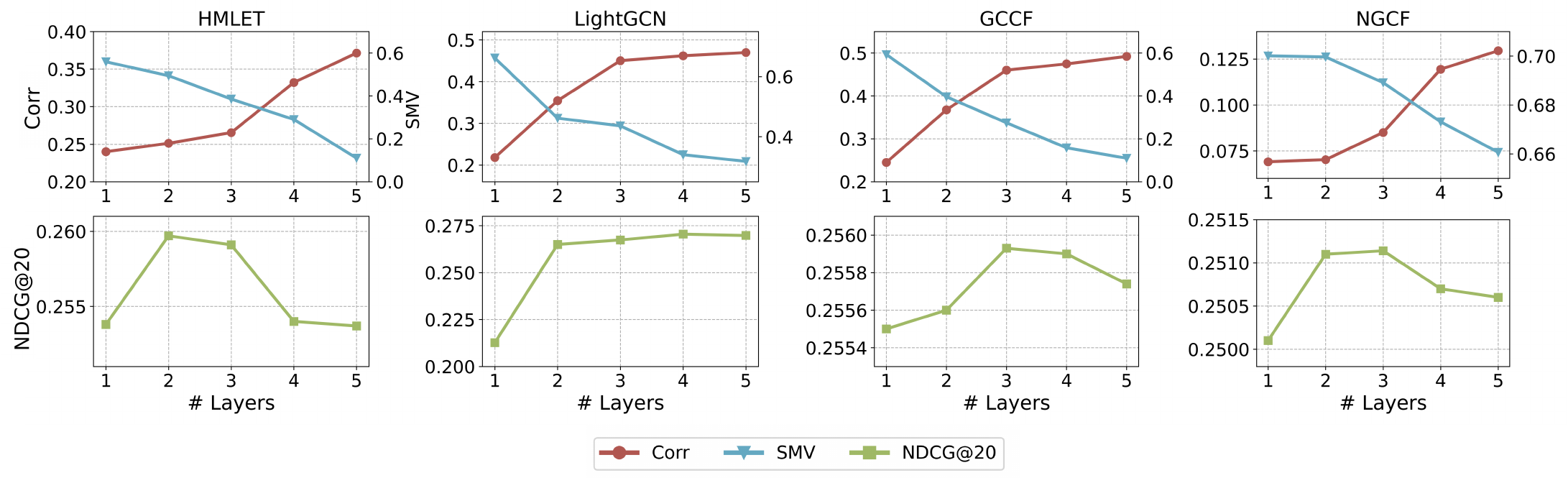}
\caption{The figures at the top illustrate the feature correlation (measured by \textit{Corr} on the left axis) and smoothness (measured by \textit{SMV} on the right axis) of representations learned by state-of-the-art GNN-based CF models on the Movielens dataset. The bottom figures present the corresponding recommendation performance (measured by NDCG@20).}
\label{fig:corrsmv}
\vspace{-2mm}
\end{figure*}

\section{Related Work}
This subsection presents pertinent literature relevant to the paper, encompassing GNN-based CF models and studies on over-correlation and over-smoothing issues.
\subsection{GNN-based Collaborative Filtering} 
Within the area of RS, GNN-based CF emerged as a forefront avenue of research. It leverages the inherent higher-order connectivity present in historical user-item interactions to attain enhanced performance compared to conventional matrix factorization techniques~\cite{mnih2007probabilistic,koren2009matrix}. During the nascent phase, \textsl{\citeauthor{li2013recommendation}} pioneered the exploration of the recommendation problem through a graph-oriented approach, treating CF as a link prediction task within bipartite graphs~\cite{li2013recommendation}. Subsequently, propelled by the progression of GNNs, researchers progressively delved into GNNs' integration within the realm of RS. As an illustration, GCMC~\cite{berg2017graph} established a graph-based auto-encoder framework employing GNNs to facilitate the augmentation of the rating matrix. SpectralCF~\cite{zheng2018spectral} directly performed spectral domain learning on the user-item bipartite graph, alleviating the cold-start problem in CF. By leveraging efficient random walks, PinSage~\cite{ying2018graph} was the first to apply GNNs to industrial-scale RS. NGCF~\cite{ngcf} injected the collaborative signal latent in the high-order connectivity between users and items into embeddings via propagation on the user-item graph. Building on this, LightGCN~\cite{he2020lightgcn} and GCCF~\cite{chen2020revisiting} subsequently streamlined the message-passing process, devising GNN paradigms tailored for CF tasks. In addition, there were also some efforts that attempted to make GNNs more suitable for CF tasks through unique designs. DGCF~\cite{wang2020disentangled}, in order to capture diverse intentions when users selected items, designed a graph disentangling module to decouple these factors and obtain disentangled representations. HMLET~\cite{kong2022linear} attempted to amalgamate linear and non-linear methodologies by applying a gating module to choose linear or non-linear propagation for each user and item. Additionally, there were other efforts aimed at expanding GNN-based recommendations by delving into aspects such as training paradigms and application.~\cite{zhang2024tgcl,wang2023setrank,wang2021personalized,wang2021variable,qin2023automatic,qin2023comprehensive}

\vspace{-4mm}

\subsection{Over-smoothing and Over-correlation} 
The proliferation of layers in GNNs gives rise to two significant challenges: over-smoothing and over-correlation. The former leads to highly similar node representations, making it difficult to distinguish between them, while the latter results in redundant information among feature representations. Both issues elevate the risk of over-fitting in GNN-based models and curtail their expressive capacity. Currently, in the field of GNNs, numerous studies have been dedicated to exploring how these two issues affect the performance of deeper GNNs in graph-related tasks~\cite{zhao2019pairnorm,liu2020towards,chen2020measuring,rusch2023survey,jin2022feature}. In the area of RS, various studies have also endeavored to mitigate the over-smoothing issue. One category of methodologies aims to streamline the process of message passing. For instance, LightGCN~\cite{he2020lightgcn} eliminated feature transformation and non-linear activation from GNNs, and GCCF~\cite{chen2020revisiting} incorporated a residual network structure while discarding non-linear activation. UltraGCN~\cite{mao2021ultragcn} skipped explicit message-passing through infinite layers. GCNII \cite{gcnii} preserved the non-linear activation in the residual network structure, and LMC \cite{lmc} scaled deep GCNII to large-scale graphs with provable convergence. The second set of approaches performs feature propagation within subdivided sub-graphs. For example, IMP-GCN~\cite{liu2021interest} employed graph convolutions within sub-graphs composed of similar-interest users and the items they interacted with. Similarly,~\citeauthor{xia2023graph} proposed a graph resizing technique that recursively divides a graph into sub-graphs. Furthermore, additional investigations~\cite{xia2022hypergraph,cai2023lightgcl,xia2023graph} endeavored to tackle the over-smoothing challenge in GNN-based CF using strategies including contrastive learning~\cite{chen2022multi} and knowledge distillation~\cite{chen2023entity}. Besides the aforementioned efforts to alleviate over-smoothing, it is noticeable that over-smoothing may not invariably be detrimental to RS. This is because the smoothness of embeddings plays a pivotal role in the effectiveness of GNN-based CF models~\cite{he2020lightgcn,fan2022graph,wang2023collaboration}.

Although significant progress has been made in the research of over-smoothing, there is currently a lack of studies on the over-correlation issue in GNN-based CF. In this paper, we analyze the impact of the over-correlation issue on GNN-based CF methods and propose a suitable solution.

\vspace{-2mm}
\section{Preliminaries}
In this section, we will introduce the paradigm of GNN-based CF.
In general, the input of CF methods consists of a user set $\mathcal{U}=\{u_1, u_2,\cdots, u_p\}$, an item set $\mathcal{I}=\{i_1, i_2,\cdots, i_q\}$, and the interaction matrix between users and items $\mathbf{R} \in \{0,1\}^{p \times q}$, where each element $r_{u,i} \in \mathbf{R}$ denotes the interaction behavior between user $u$ and item $i$, such as a purchase or click, with $1$ indicating an interaction has occurred and $0$ indicating no interaction. In GNN-based CF, the interaction matrix above is reformulated into a user-item graph $\mathcal{G}=<\mathcal{U}\cup \mathcal{I},\mathbf{A}>$, where $\mathbf{A}$ represents the adjacency matrix: \begin{equation} \small 
\mathbf{A}=\begin{bmatrix}\mathbf{0}&\mathbf{R}\\\mathbf{R}^\mathrm{T}&\mathbf{0}\end{bmatrix}.
  \label{eq:a}
\end{equation}
Aggregation operations and update operations can represent the message-passing process on the user-item graph~\cite{wu2022graph}: \begin{equation} \small
    \begin{split}&\mathbf{e}_i^{(l+1)}=\mathrm{Updater}\left(\mathbf{e}_i^{(l)},\mathrm{Aggregator}\left(\left\{\mathbf{e}_u^l,\forall u\in\mathcal{N}_i\right\}\right)\right)\\&\mathbf{e}_u^{(l+1)}=\mathrm{Updater}\left(\mathbf{e}_u^{(l)},\mathrm{Aggregator}\left(\left\{\mathbf{e}_i^l,\forall i\in\mathcal{N}_u\right\}\right)\right)\end{split},
\label{eq:mp} 
\end{equation}where $\mathbf{e}_u^{(l)},\mathbf{e}_i^{(l)}$ represents the representation of users or items at layer $l$. $\mathcal{N}_u,\mathcal{N}_i$ represent the neighboring nodes of user $u$ or item $i$, respectively. Aggregator denotes the function responsible for aggregation operations, such as mean-pooling and attention mechanisms. Updater refers to the operation function that updates the current node representation, for instance, sum operation and nonlinear transformation. 

After the message-passing process, node representation matrix:\begin{equation} \small
    \mathbf{E}^{(l)}=\left[\mathbf{e}_{u_{1}}^{(l)}, \cdots, \mathbf{e}_{u_{p}}^{(l)}, \mathbf{e}_{i_{1}}^{(l)}, \cdots, \mathbf{e}_{i_{q}}^{(l)}\right]
\label{eq:mp}
\end{equation} can be obtained at each layer (as shown in
Figure~\ref{fig:illu}). The final node representation matrix, denoted by $\mathbf{E}$, can be obtained through various pooling operations, such as mean/sum/weighted-pooling and concatenation: \begin{equation} \small
   \mathbf{E} = \mathrm{Pooling}(\mathbf{E}^{(0)},\mathbf{E}^{(1)},\cdots,\mathbf{E}^{(L)}),
\label{eq:pooling} 
\end{equation} where $L$ denotes the number of layers, $\mathbf{E}^{(0)}$ represents learnable embedding matrix.
Without losing generalization, this paper's analysis will focus on LightGCN~\cite{he2020lightgcn}, one of the most representative methods in GNN-based CF. Its message-passing process can be formulated as follows: \begin{equation} \small
   \mathbf{E}^{(l+1)}=(\mathbf{D}^{-\frac12}\mathbf{A}\mathbf{D}^{-\frac12})\mathbf{E}^{(l)}, 
\label{eq:lgn}
\end{equation} where $\mathbf{D}$ is the degree matrix of $\mathbf{A}$.

\begin{figure*}[!t]
\centering\includegraphics[width=0.78\textwidth]{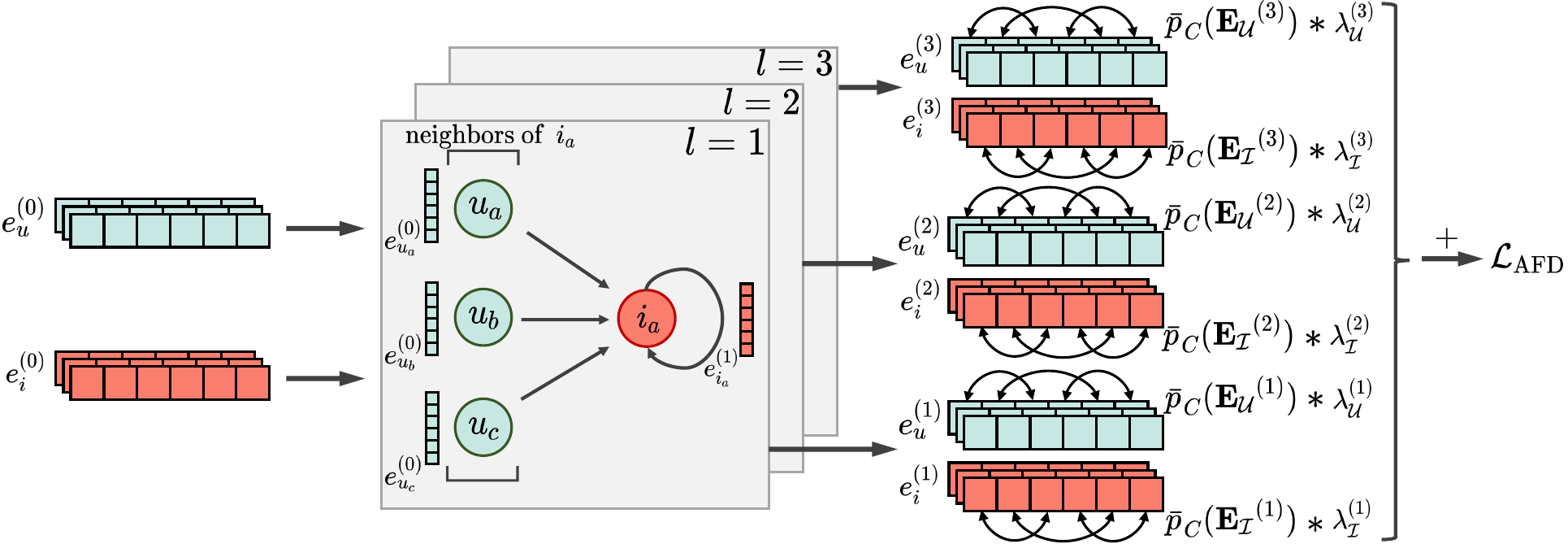}
\caption{An overview of the AFDGCF framework, highlighting its model-agnostic nature and dynamic application of correlation penalties to user and item representations at each layer for enhanced performance.}
\label{fig:AFDGCF}
\end{figure*}
\section{The General AFDGCF Framework}

In this section, we commence with an empirical analysis of the impact of over-correlation and over-smoothing issues on GNN-based CF models. Subsequently, we delve into their theoretical connection. Building upon this, we propose an Adaptive Feature De-correlation Graph Collaborative Filtering (AFDGCF) framework, as illustrated in Figure~\ref{fig:AFDGCF}, which addresses the issues of over-correlation and over-smoothing in GNN-based CF simultaneously while finding a trade-off between over-smoothing and representation smoothness through layer-wise adaptive feature correlation constraint.

\subsection{Empirical Analysis on Over-correlation and Over-smoothing}\label{SEC:Theo}
Over-correlation issue refers to the increasing correlation between the feature dimensions of the GNN model's output representations as the number of GNN layers increases~\cite{chen2020mmea, sun2023hierarchical, chen2024collaboration}. Over-smoothing issue refers to the growing similarity between the node representations of the GNN model's output as the number of GNN layers increases. The former will lead to feature redundancy among the learned representations, while the latter will lead to convergence of node representations, making them difficult to be distinguished. In recommender systems, both of them will constrain the quality of learned user and item representations, increase the risk of over-fitting, and result in sub-optimal model performance. To investigate their impact on GNN-based CF models, we adopt \textit{Corr}~\cite{jin2022feature} and \textit{SMV}~\cite{liu2020towards}  to assess them respectively.

\textit{Corr} is utilized to measure the correlation between feature dimensions, which calculates the average Pearson correlation coefficient~\cite{cohen2009pearson} between any two feature dimensions: \begin{equation} \small
   \textit{Corr}(\mathbf{E})=\frac1{d(d-1)}\sum_{i\neq j}|\rho(\mathbf{E}_{\ast i},\mathbf{E}_{\ast j})|,\quad i,j\in\{1,2,\ldots,d\},
\label{eq:corr} 
\end{equation} where $d$ represents the dimension of the features, $\mathbf{E}_{\ast i}$ denotes the $i$-th feature of users and items, and $\rho(\mathbf{x},\mathbf{y})$ represents the Pearson correlation coefficient between vectors $\mathbf{x}$ and $\mathbf{y}$:
\begin{equation} \small
   \rho(\mathbf{x},\mathbf{y})=\frac{\sum\left(x_i-\bar{x}\right)\left(y_i-\bar{y}\right)}{\sqrt{\sum\left(x_i-\bar{x}\right)^2\sum\left(y_i-\bar{y}\right)^2}},
\label{eq:rho}
\end{equation}
which ranging from $-1$ to $1$. The larger the absolute value, the stronger the correlation between the corresponding vectors.

\textit{SMV} is utilized to gauge the similarity between user and item representations, which computes the average normalized Euclidean distance between any two nodes: \begin{equation} \small
\begin{split} 
    \textit{SMV}(\mathbf{E})=\frac{1}{m(m-1)}\sum_{i,j \in \mathcal{U}\cup\mathcal{I}, i \neq j} D\left(\mathbf{E}_{i \ast},\mathbf{E}_{j \ast}\right),
\end{split}
\label{eq:smv}
\end{equation} where $D(\mathbf{x},\mathbf{y})$ denotes the normalized euclidean distance between any two vectors $\mathbf{x}$ and $\mathbf{y}$:
\begin{equation} \small
   D(\mathbf{x},\mathbf{y})=\frac12\left\|\frac{\mathbf{x}}{\|\mathbf{x}\|}-\frac{\mathbf{y}}{\|\mathbf{y}\|}\right\|.
\label{eq:d}
\end{equation}

Based on the two metrics above, we explore over-correlation and over-smoothing issues in GNN-based CF models. From Figure~\ref{fig:corrsmv}, we observe the following trends for these four models: 
\begin{itemize}
    \item \textit{Corr} increases with layer count, indicating an increase in the correlation among the learned representations. 
    \item \textit{SMV} decreases with layer count, indicating an increase in the smoothness among the learned representations. 
    \item The recommendation performance does not consistently improve with the increase in the number of layers; instead, performance degradation occurs after reaching 2 to 3 layers. 
\end{itemize}
The above three points reveal the widespread existence of over-correlation and over-smoothing issues in GNN-based CF models, leading to the suboptimal performance of the models. Moreover, the severity of both issues increases with the number of layers, indicating a certain association between them. Next, we will analyze their association theoretically and propose solutions.

\subsection{Theoretical Analysis on Over-correlation and Over-smoothing}
In this subsection, we endeavor to analyze the relationship between over-correlation and over-smoothing theoretically. Firstly, we define the concepts of column correlation and row correlation of the representation matrix, which correspond to the feature correlation and smoothness, respectively. Then, through mathematical proof, we demonstrate a proportional relationship between the column correlation and row correlation of the matrix, thus inferring the connection between over-correlation and over-smoothing issues.

In general, when calculating row or column correlation coefficients, it is common practice to first normalize the matrix to have a mean of $0$ and a variance of $1$~\cite{cohen2009pearson}. Here, to simplify the analysis process furthermore, we assume that the representation matrix $\mathbf{E}$ satisfies double standardization, \textit{i.e.}, each row and column of $\mathbf{E}$ have a mean of $0$ and a variance of $1$. Under such assumptions, we can define the column correlation coefficient matrix of $\mathbf{E}$ as \begin{equation} \small
    \mathbf{P_C} = \frac{1}{m} \mathbf{E}^T\mathbf{E}\;, \quad \mathbf{P_C}_{ij} = \rho(\mathbf{E}_{\ast i},\mathbf{E}_{\ast j}).
    \label{corr_matrix_C}
\end{equation} Similarly, the row correlation coefficient matrix of $\mathbf{E}$ is 
\begin{equation} \small 
    \mathbf{P_R} = \frac{1}{n}\mathbf{E}\mathbf{E}^T, \quad \mathbf{P_R}_{ij} = \rho(\mathbf{E}_{i \ast},\mathbf{E}_{j \ast}).
    \label{corr_matrix_R}
\end{equation} Based on this, we can use $\|\mathbf{P_C}\|_F$ and $\|\mathbf{P_R}\|_F$ to describe the column correlation and row correlation of matrix $\mathbf{E}$, respectively. In fact, the relationship between row correlation and column correlation of a matrix has been studied in previous research~\cite{efron2008row}. Next, we will demonstrate the association between them based on this.

\textsc{Theorem 1.} \textit{If matrix $\mathbf{E}$ satisfies double standardization, then its corresponding row correlation matrix $\mathbf{P_R}$ and column correlation matrix $\mathbf{P_C}$ satisfy} $\|\mathbf{P_R}\|_F \propto \|\mathbf{P_C}\|_F$.

\begin{proof}
The singular value decomposition of matrix $\mathbf{E}$ is: \begin{equation} \small
    \underset{\strut m \times n}{\mathbf{E}}=\underset{\strut m \times d}{\mathbf{U}} \;\underset{\strut d \times d}{\mathbf{\Sigma}} \;\underset{\strut d \times n}{\mathbf{V}^T},
\label{eq:svd}
\end{equation} where $\mathbf{\Sigma}$ is the diagonal matrix of ordered singular values, $\mathbf{U}$ and $\mathbf{V}$ are orthogonal matrices satisfying \begin{equation} \small
\mathbf{U}^T\mathbf{U}=\mathbf{V}^T\mathbf{V}=\mathbf{I}_{d},
\end{equation} where $\mathbf{I}_{d}$ is the $d\times d$ identity. The squares of the diagonal elements: \begin{equation} \small
   e_1\geq e_2\geq\cdots\geq e_d>0 \quad(e_i=\mathbf{\Sigma}_i^2),
\label{eq:e}
\end{equation} are the eigenvalues of \begin{equation} \small
    \mathbf{E}^T\mathbf{E}=\mathbf{V}^T\mathbf{\Sigma}^{2}\mathbf{V}.
\end{equation} Then we have: \begin{equation} \small
    \begin{aligned}&\frac{\sum_{i=1}^n\sum_{j=1}^n{\mathbf{P_C}}_{ij}^2}{n^2}=\frac{\operatorname{tr}((\mathbf{E}^{T}\mathbf{E})^2)}{m^2n^2}=\frac{\operatorname{tr}(\mathbf{V}^{T}\mathbf{\Sigma}^4\mathbf{V})}{m^2n^2}=c,\\
    &\frac{\sum_{i=1}^n\sum_{j=1}^n{\mathbf{P_R}}_{ij}^2}{m^2}=\frac{\operatorname{tr}((\mathbf{E}\mathbf{E}^{T})^2)}{m^2n^2}=\frac{\operatorname{tr}(\mathbf{U}\mathbf{\Sigma}^4\mathbf{U}^{T})}{m^2n^2}=c,\end{aligned}
\label{eq:proof}
\end{equation} where $tr(\cdot)$ represents the trace of a matrix and $c$ can be computed to be a constant 
\begin{equation} \small
    c=\sum_{i=1}^d\frac{e_i^2}{(mn)^2}.
\end{equation} That is to say: 
\begin{equation} \small 
    \frac{1}{m}\|\mathbf{P_R}\|_F = \frac{1}{n}\|\mathbf{P_C}\|_F.
\end{equation}
\end{proof}
\vspace{-3mm}

The above theory reveals a proportional relationship between column correlation and row correlation of a matrix. In the GNN-based CF model, for the representation matrix $\mathbf{E}$, the column correlation represents the correlation between features, and the row correlation describes the similarity between user and item representations, which can be considered as a proxy for measuring smoothness. The larger the row correlation, the smaller the average difference between representations, indicating a higher level of smoothing. Therefore, we can conclude that over-correlation and over-smoothing positively correlate in degree. This is consistent with the observations from previous experiments. According to this conclusion, we can utilize the following proposed method to address the issue of over-correlation while mitigating the impact of over-smoothing.

\subsection{Feature De-correlation}
To alleviate the excessively high correlation in feature dimensions of user and item representations learned by GNN-based CF models, a straightforward approach is to design a loss that constrains the correlation between features. In terms of the types of loss, we have a variety of choices. If we consider each column of the representation matrix as a high-dimensional vector, we can impose constraints between any two columns using different distance metrics (e.g., Euclidean distance or cosine distance); if we view each dimension of the feature as a random variable, we can impose penalties based on their KL-divergence or correlation coefficient. Considering the substantial computational effort required to calculate the sum of KL-divergence between any two columns of a matrix, and the fact that Euclidean distance, cosine distance, and Pearson correlation coefficient become equivalent after data standardization, we naturally opt to design our loss function based on the correlation coefficient.

Our objective is to reduce the average correlation between columns of the representation matrix $\mathbf{E}$. Formally, \begin{equation} \small
\bar{p}_C(\mathbf{E}) = \frac{1}{d(d-1)}\sum_{i\neq j}\rho(\mathbf{E}_{\ast i},\mathbf{E}_{\ast j})^2, \quad i,j\in\{1,2,\ldots,d\}.
\label{eq:aim} 
\end{equation} Since the correlation coefficient of a vector with itself is always $1$, we can also represent it using the previously defined column correlation coefficient matrix $\mathbf{P_R}$: \begin{equation} \small 
\Bar{p}_C(\mathbf{E}) = \sqrt[]{ \frac{1}{2} \|\mathbf{P_C}-\mathbf{I}_{d}\|_F^2} = \frac{1}{\sqrt[]{2} } \|\mathbf{P_C}-\mathbf{I}_{d}\|,
\label{eq:aimmatrix} 
\end{equation}where $\mathbf{I}_{d}$ is the $d\times d$ identity and $\mathbf{P_R}$ can be calculated using the following formula: \begin{equation} \small
\mathbf{P_C}_{ij}=\frac{\mathbf{Cov_C}_{ij}}{\sqrt{\mathbf{Cov_C}_{ii} \, \mathbf{Cov_C}_{jj}}},
\end{equation} where $\mathbf{Cov_C}$ denotes the covariance matrix among the column vectors of the representation matrix $\mathbf{E}$: \begin{equation} \small 
    \mathbf{Cov_C} = (\mathbf{E}-\bar{\mathbf{E}})^T(\mathbf{E}-\bar{\mathbf{E}}), \label{eq:covc} 
\end{equation} where $\bar{\mathbf{E}}$ represents a matrix composed of the mean values of each column of $\bar{\mathbf{E}}$. It is worth noting that here we utilize the entire representation matrix to compute the correlations between feature dimensions. In fact, to reduce the computational complexity, when dealing with large-scale user and item sets, it is possible to randomly sample a batch of users and items for calculating the correlations between feature dimensions.
\subsection{Adaptive Feature De-correlation}
Directly applying the function $\Bar{p}_C(\cdot)$ to the learned representation matrix $\mathbf{E}$ of the GNN-based CF model is not the optimal choice. One critical consideration is that various GNN-based CF models employ distinct pooling strategies. For example, GCCF uses concatenation, while LightGCN uses mean pooling. This results in significant differences in the correlation of the representation matrix $\mathbf{E}$ for different models. To address this and ensure our approach remains model-agnostic, we propose penalizing the representations obtained after each message-passing operation as an alternative approach. Furthermore, since user and item feature distributions often exhibit substantial differences, applying penalties separately to the user and item representation matrix is necessary. This approach also helps in preserving the relationships between users and items learned by the model, thus avoiding potential disruptions. In this case, our feature de-correlation loss can be formulated in the following manner:  \begin{equation} \small
    \mathcal{L}_\text{AFD}= \sum_{l=1}^L \lambda_\mathcal{U}^{(l)} \bar{p}_C(\mathbf{E_\mathcal{U} }^{(l)}) + \lambda_\mathcal{I}^{(l)}\bar{p}_C(\mathbf{E_\mathcal{I} }^{(l)}).
\end{equation} It is worth noting that we do not impose a penalty on $\mathbf{E_\mathcal{U} }^{(0)},\mathbf{E_\mathcal{I} }^{(0)}$ because the $0$-th layer represents the initial user and item embeddings without undergoing graph convolution, usually subject to $L_2$ regularization constraints. ${\lambda_\mathcal{U}}^{(l)},{\lambda_\mathcal{I}}^{(l)}$ represents the penalty coefficient for the $l$-th layer corresponding to users and items. If fixed at $1/L$, it indicates a fixed penalty of the same magnitude for each layer. However, this is not the most reasonable approach. The strength of penalty accepted between different layers should be dynamically adjusted during training based on the relative correlation magnitudes of each layer.

\begin{table}[t]
\caption{Statistics of the utilized datasets.}
\vspace{-2mm}
\label{table:datasets}
\centering
\begin{tabular}{ccccc}
  \toprule
  \textbf{Datasets} & \textbf{\# User} & \textbf{\# Item} & \textbf{\# Interaction} & \textbf{Density} \\
  \midrule
  Movielens & 6,040 & 3,629 & 836,478 & $3.8e^{-2}$ \\
  Yelp & 26,752 & 19,857 & 1,001,901 & $1.8e^{-3}$ \\
  Gowalla & 29,859 & 40,989 & 1,027,464 & $8.4e^{-4}$ \\
  Amazon-book & 58,145 & 58,052 & 2,517,437 & $7.5e^{-4}$ \\
  \bottomrule
\end{tabular}
\vspace{-3mm}
\end{table}

For a GNN-based CF model with a specific number of layers, the representations of deeper layers often have greater embedding smoothness while having higher feature correlations. Considering the contribution of feature smoothness to the effectiveness of GNN-based CF models~\cite{he2020lightgcn,fan2022graph,wang2023collaboration}, it is crucial to maintain the smoothness of the representations of deeper layers while restricting the feature correlations of the model's final representations. Therefore, we propose the strategies which allocates lower penalty coefficients to the representations of deeper layers and maintains the total penalty amount, as follows: \begin{equation} \small
{\lambda_\mathcal{U}}^{(l)} = \frac{1/\bar{p}_C(\mathbf{E_\mathcal{U}}^{(l)})}{\sum_{i=1}^L 1/\bar{p}_C(\mathbf{E_\mathcal{U}}^{(i)})},\; {\lambda_\mathcal{I}}^{(l)} = \frac{1/\bar{p}_C(\mathbf{E_\mathcal{I}}^{(l)})}{\sum_{i=1}^L 1/\bar{p}_C(\mathbf{E_\mathcal{I}}^{(i)})}.
\label{eq:covr}
\end{equation} If the adaptive strategy is not employed, throughout the entire training process, the penalty coefficients for user and item correspondence at each layer remain fixed and unchanging. However, with the adoption of the adaptive strategy, the penalty coefficients for user and item correspondence at each layer will dynamically change in each training step and tend to stabilize as the model approaches convergence. The final loss function is shown below: \begin{equation} \small
    \mathcal{L} = \mathcal{L}_\text{CF} + \alpha \mathcal{L}_\text{AFD},
    \label{eq:overall_loss}
\end{equation} where $\mathcal{L}_\text{CF}$ represents the original loss functions of the CF model like Bayesian Personalized Ranking (BPR) loss~\cite{rendle2012bpr}, Cross-Entropy (CE) loss, etc. $\alpha$ is a hyper-parameter used to adjust the relative magnitude of the adaptive feature de-correlation loss.

\begin{table*}[!t]
\caption{Performance comparisons on four datasets. We have conducted the paired \textit{t}-test to verify that the difference between each base model and our proposed method is statistically significant for $p < 0.05$. (Underlines represent the optimal performance)}
\vspace{-2mm}
\label{table:compar}
\centering
\renewcommand\arraystretch{0.9}
\begin{tabular}{c|ccc|ccc|ccc|ccc}
\hline 
\textbf{Dataset} & \multicolumn{3}{c|}{\textbf{MovieLens}}  & \multicolumn{3}{c|}{\textbf{Yelp}} & \multicolumn{3}{c|}{\textbf{Gowalla}} & \multicolumn{3}{c}{\textbf{Amazon-book}}  \\ \hline
\textbf{Method} & \textbf{Recall} & \textbf{NDCG} & \textbf{MAP}  & \textbf{Recall}  & \textbf{NDCG}  & \textbf{MAP} & \textbf{Recall} & \textbf{NDCG} & \textbf{MAP} & \textbf{Recall} & \textbf{NDCG} & \textbf{MAP} \\ \hline \hline
BPRMF & 0.1697 & 0.2353 & 0.1347 & 0.0643  & 0.0437  & 0.0240  & 0.1110 & 0.0790 & 0.0509  & 0.0678 & 0.0470 & 0.0273  \\
DMF  & 0.1739 & 0.2338 & 0.1320 & 0.0560  & 0.0375  & 0.0207  & 0.0828 & 0.0589 & 0.0371  & 0.0643 & 0.0447 & 0.0260  \\
ENMF & 0.1941 & 0.2613 & 0.1524 & 0.0742  & 0.0520  & 0.0298  & 0.1271 & 0.0914 & 0.0597 & 0.0839 & 0.0617 & 0.0374  \\
MultiVAE & 0.1889 & 0.2370 & 0.1318 & 0.0731  & 0.0494  & 0.0275  & 0.1188 & 0.0838 & 0.0550  & 0.0816 & 0.0574 & 0.0352  \\
RecVAE & 0.1852 & 0.2524 & 0.1475 & 0.0777  & 0.0536  & 0.0301  & 0.1466 & 0.1046 & 0.0695  & 0.1044 & 0.0772 & 0.0484  \\ 
DGCF & 0.1819 & 0.2477 & 0.1429 & 0.0712 & 0.0487 & 0.0272 & 0.1372 & 0.0974 & 0.0635 & 0.0951 & 0.0689 & 0.0424  \\ \hline
NGCF & 0.1743 & 0.2404 & 0.1376 & 0.0613  & 0.0419  & 0.0229  & 0.1129 & 0.0800 & 0.0519  & 0.0695 & 0.0484 & 0.0287  \\
\textbf{AFD-NGCF} & \textbf{0.1764} & \textbf{0.2449} & \textbf{0.1416} & \textbf{0.0636}  & \textbf{0.0432}  & \textbf{0.0237}  & \textbf{0.1233} & \textbf{0.0869} & \textbf{0.0566}  & \textbf{0.0735} & \textbf{0.0511} & \textbf{0.0304}  \\
\textbf{Improv.} & \textbf{1.20\%} & \textbf{1.87\%} & \textbf{2.91\%} & \textbf{3.75\%}  & \textbf{3.10\%}  & \textbf{3.49\%}  & \textbf{9.21\%} & \textbf{8.63\%} & \textbf{9.05\%}  & \textbf{5.76\%} & \textbf{5.58\%} & \textbf{5.92\%}  \\ \hline
GCCF & 0.1761 & 0.2460 & 0.1431 & 0.0658  & 0.0451  & 0.0247  & 0.1268 & 0.0907 & 0.0589  & 0.0898 & 0.0639 & 0.0387  \\
\textbf{AFD-GCCF} & \textbf{0.1882} & \textbf{0.2566} & \textbf{0.1505} & \textbf{0.0739}  & \textbf{0.0506}  & \textbf{0.0284}  & \textbf{0.1416} & \textbf{0.1015} & \textbf{0.0667}  & \textbf{0.0977} & \textbf{0.0698} & \textbf{0.0427}  \\
\textbf{Improv.} & \textbf{6.87\%} & \textbf{4.31\%} & \textbf{5.17\%} & \textbf{12.31\%} & \textbf{12.20\%} & \textbf{14.98\%} & \textbf{11.67\%} & \textbf{11.91\%} & \textbf{13.24\%}  & \textbf{8.80\%} & \textbf{9.23\%} & \textbf{10.34\%} \\ \hline
HMLET & 0.1799 & 0.2479 & 0.1433 & 0.0723  & 0.0503  & 0.0285  & 0.1427 & 0.1028 & 0.0679  & 0.0967 & 0.0699 & 0.0430  \\
\textbf{AFD-HMLET}  & \textbf{0.1922} & \textbf{0.2606} & \textbf{0.1523} & \textbf{0.0810} & \textbf{0.0562}  & \textbf{0.0319}  & \textbf{0.1541} & \textbf{0.1101} & \textbf{0.0727}  & \textbf{0.1039} & \textbf{0.0760}  & \textbf{0.0471}  \\
\textbf{Improv.} & \textbf{6.84\%} & \textbf{5.12\%} & \textbf{6.28\%} & \textbf{12.03\%} & \textbf{11.73\%} & \textbf{11.93\%} & \textbf{7.99\%} & \textbf{7.10\%} & \textbf{7.07\%}  & \textbf{7.45\%} & \textbf{8.73\%} & \textbf{9.53\%}  \\ \hline
LightGCN  & 0.1886 & 0.2540 & 0.1470 & 0.0756  & 0.0525  & 0.0295  & 0.1433 & 0.1019 & 0.0664  & 0.1015 & 0.0744 & 0.0465  \\
\textbf{AFD-LightGCN} & \underline{\textbf{0.1985}} & \underline{\textbf{0.2689}} & \underline{\textbf{0.1594}} & \underline{\textbf{0.0831}}  & \underline{\textbf{0.0575}}  & \underline{\textbf{0.0327}}  & \underline{\textbf{0.1564}} & \underline{\textbf{0.1117}} & \underline{\textbf{0.0736}}  & \underline{\textbf{0.1078}} & \underline{\textbf{0.0781}} & \underline{\textbf{0.0486}}  \\
\textbf{Improv.} & \textbf{5.25\%} & \textbf{5.87\%} & \textbf{8.44\%} & \textbf{9.92\%}  & \textbf{9.52\%}  & \textbf{10.85\%} & \textbf{9.14\%} & \textbf{9.62\%} & \textbf{10.84\%} & \textbf{6.21\%} & \textbf{4.97\%} & \textbf{4.52\%}  \\ \hline
\end{tabular}
\vspace{-3mm}
\end{table*}

Overall, the proposed AFDGCF framework is a general solution specifically targeting the issue of feature over-correlation in GNN-based CF models. This framework is model-agnostic and can be effortlessly combined with a wide variety of GNN-based CF models, such as NGCF~\cite{wang2019neural}, GCCF~\cite{chen2020revisiting}, LightGCN~\cite{he2020lightgcn}, HMLET~\cite{kong2022linear}, etc. This integration is achieved by incorporating the loss $\mathcal{L}_\text{AFD}$ into the existing CF loss function, as detailed in Equation~\ref{eq:overall_loss}.
Furthermore, the AFDGCF framework is also helpful in addressing the over-smoothing issue. This dual capability is underpinned by our theoretical analysis, which examines the intricate dynamics between these two prevalent problems in GNN-based CF models.
\section{Experiments}

To better understand the capabilities and effectiveness of our proposed AFDGCF framework, we instantiated it on four representative GNN-based CF models and evaluated its performance on four publicly available datasets. Specifically, we will answer the following research questions to unfold the experiments:  \textbf{RQ1}: What degree of performance improvement can our proposed AFDGCF framework bring to the state-of-the-art methods in GNN-based CF? \textbf{RQ2}: Does adaptive de-correlation have a performance advantage compared to fixed one? \textbf{RQ3}: How do the hyper-parameters influence the effectiveness of the proposed AFDGCF framework? \textbf{RQ4}: What impact does the AFDGCF framework have on the correlation of the learned representations?

\begin{figure*}[!t]
\centering\includegraphics[width=0.85\textwidth]{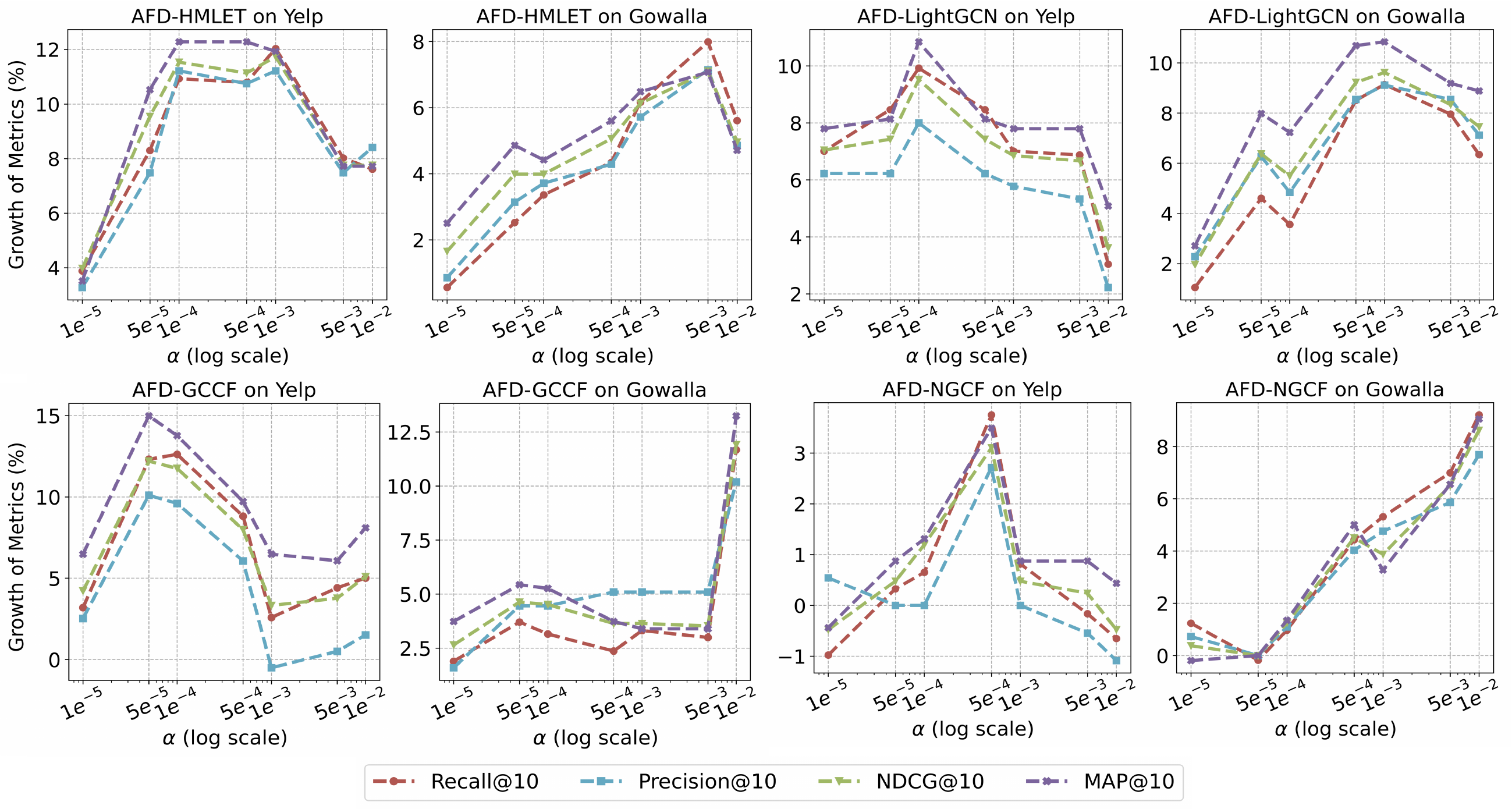}
\caption{Performance comparison \textit{w.r.t.} different $\alpha$. Using the results of the original model (\textit{i.e.}, $\alpha=0$) as the reference.}
\label{fig:hpyer}
\end{figure*}

\begin{table}[t]
\caption{Comparison of training efficiency on four datasets.}
\vspace{-2mm}
\label{table:effi}
\centering
\renewcommand\arraystretch{1.05}
\resizebox{\columnwidth}{!}{
\begin{tabular}{c|c|c|c|c}
\hline
\textbf{Dataset} & \textbf{HMLET} & \textbf{AFD-H} & \textbf{LightGCN} & \textbf{AFD-L} \\ \hline \hline
MovieLens   & \begin{tabular}[c]{@{}c@{}}167  eps\\ 5.95s/ep\end{tabular}  & \begin{tabular}[c]{@{}c@{}}112  eps\\ 6.87s/ep\end{tabular}  & \begin{tabular}[c]{@{}c@{}}309  eps\\ 4.51s/ep\end{tabular}  & \begin{tabular}[c]{@{}c@{}}114  eps\\ 5.12s/ep\end{tabular}  \\ \hline
Yelp  & \begin{tabular}[c]{@{}c@{}}114  eps\\ 9.54s/ep\end{tabular}  & \begin{tabular}[c]{@{}c@{}}109  eps\\ 10.58s/ep\end{tabular} & \begin{tabular}[c]{@{}c@{}}364  eps\\ 6.76s/ep\end{tabular}  & \begin{tabular}[c]{@{}c@{}}151  eps\\ 7.31s/ep\end{tabular}  \\ \hline
Gowalla     & \begin{tabular}[c]{@{}c@{}}347  eps\\ 10.25s/ep\end{tabular} & \begin{tabular}[c]{@{}c@{}}205  eps\\ 11.27s/ep\end{tabular}  & \begin{tabular}[c]{@{}c@{}}396  eps\\ 6.98s/ep\end{tabular}  & \begin{tabular}[c]{@{}c@{}}302  eps\\ 7.82s/ep\end{tabular}  \\ \hline
Amazon-book & \begin{tabular}[c]{@{}c@{}}458  eps\\ 58.12s/ep\end{tabular} & \begin{tabular}[c]{@{}c@{}}198  eps\\ 61.23s/ep\end{tabular}  & \begin{tabular}[c]{@{}c@{}}742  eps\\ 32.06s/ep\end{tabular} & \begin{tabular}[c]{@{}c@{}}360  eps\\ 35.73s/ep\end{tabular} \\ \hline
\end{tabular}
}
\end{table}

\subsection{Experimental Settings}

\textbf{Datasets and metrics.} We selected four publicly available datasets to evaluate the performance of the proposed AFDGCF, including MovieLens~\cite{harper2015movielens}, Yelp~\footnote{https://www.yelp.com/dataset}, Gowalla~\cite{cho2011friendship}, and Amazon-book~\cite{mcauley2015image}, which vary in domain, scale, and density. The statistical information for these four datasets is summarized in Table~\ref{table:datasets}. For the Yelp and Amazon-book datasets, we excluded users and items with fewer than 15 interactions to ensure data quality. Similarly, for the Gowalla dataset, we removed those with fewer than 10 interactions. Moreover, we partitioned each dataset into training, validation, and testing sets using an 8:1:1 ratio. We chose Recall@K, Normalized Discounted Cumulative Gain (NDCG)@K, and Mean Average Precision (MAP)@K as the evaluation metrics (K=10). We also employed the all-ranking protocol~\cite{he2020lightgcn} to avoid bias arising from sampling.

\noindent
\textbf{Methods for Comparison.}
To verify the effectiveness of our proposed method, we compared it with several state-of-the-art models: BPRMF~\cite{rendle2012bpr}, DMF~\cite{xue2017deep}, ENMF~\cite{chen2020efficient}, MultiVAE~\cite{liang2018variational}, RecVAE~\cite{shenbin2020recvae}, DGCF~\cite{wang2020disentangled}, NGCF~\cite{wang2019neural}, GCCF~\cite{chen2020revisiting}, LightGCN~\cite{he2020lightgcn}, HMLET~\cite{kong2022linear}. AFD-NGCF/GCCF/LightGCN/HMLET are the instantiation of the AFDGCF framework we propose on those above four GNN-based CF models. Abbreviated as AFD-N/G/L/H.

\begin{table}[t]
\caption{Ablation study of AFDGCF.}
\vspace{-2mm}
\label{table:abla}
\centering
\renewcommand\arraystretch{1.05}
\resizebox{\columnwidth}{!}{
\begin{tabular}{cc|c|c|c|c}
\hline
\textbf{Dataset} & \textbf{Metric} & \textbf{\;AFD-N\;}  & \textbf{AFD-f-N}  & \textbf{\;AFD-G\;}  & \textbf{AFD-f-G} \\ \hline \hline
\multirow{2}{*}{Yelp}  & Recall & \textbf{0.0636} & 0.0623  & \textbf{0.0739} & 0.0719 \\
 & NDCG & \textbf{0.0432} & 0.0424  & \textbf{0.0506} & 0.0487 \\ \hline
\multirow{2}{*}{Gowalla} & Recall & \textbf{0.1233} & 0.1198  & \textbf{0.1315} & 0.1298 \\
 & NDCG & \textbf{0.0869} & 0.0850  & \textbf{0.0949} & 0.0937 \\ \hline \\ \hline
\textbf{Dataset} & \textbf{Metric} & \textbf{AFD-H} & \textbf{AFD-f-H} & \textbf{AFD-L} & \textbf{AFD-f-L} \\ \hline \hline
\multirow{2}{*}{Yelp}  & Recall & \textbf{0.0810} & 0.0796  & \textbf{0.0831} & 0.0808 \\
 & NDCG & \textbf{0.0562} & 0.0552  & \textbf{0.0575} & 0.0558 \\ \hline
\multirow{2}{*}{Gowalla} & Recall & \textbf{0.1541} & 0.1465  & \textbf{0.1564} & 0.1505 \\
 & NDCG & \textbf{0.1101} & 0.1069  & \textbf{0.1117} & 0.1081 \\ \hline
\end{tabular}
}
\vspace{-5mm}
\end{table}

\noindent
\textbf{Implementation Details.} 
For a fair comparison, we employed the Adam optimizer ~\cite{kingma2014adam} for all models with a learning rate of $1e^{-3}$ and set the training batch size to $4096$. The embedding size or hidden dimension in all models were set to $128$, and model parameters were initialized using the Xavier distribution. Furthermore, we used early stopping for the indicator NDCG@10, halting training when there was no further increase in NDCG@10 on valid set to prevent the model from over-fitting. For the four GNN-based CF models, we performed 3-layer propagation. As for the two VAE-based models, we set their encoder and decoder as a 1-hidden-layer MLP with $[n, 600, 128, 600, n]$ dimensions. The weight of L2 regularization is selected within the range $\{1e^{-2},1e^{-3}, 1e^{-4}, 1e^{-5},1e^{-6}\}$, and dropout ratios were selected from $\{0.0, 0.1, \cdots, 0.8, 0.9\}$. Other model-specific hyper-parameters were configured according to the settings prescribed in the original papers. For the proposed AFDGCF framework, we tuned the hyper-parameter $\alpha$ within the range $\{1e^{-5}, 5e^{-5}, \cdots, 5e^{-2}, 1e^{-1}\}$. The implementation of all models above and their evaluation was conducted using the open-source framework RecBole ~\cite{recbole[1.0]}.

\vspace{-1mm}

\subsection{Performance Comparisons (RQ1)}
Table~\ref{table:compar} demonstrates the effectiveness of our proposed AFDGCF framework as instantiated on four GNN-based CF models and provides a comparative analysis against various baseline methods across four distinct datasets. The table reveals that implementing AFDGCF in all the aforementioned GNN-based CF models is effective. This stems from notable performance enhancements achieved by mitigating the over-correlation issue, while successfully balancing mitigating over-smoothing and maintaining embedding smoothness. Furthermore, the AFDGCF framework demonstrates even more pronounced performance enhancements on large-scale datasets. On Yelp, for Recall@10, ADFGCF achieves performance improvements of $3.75\%$, $12.31\%$, $12.03\%$, and $9.92\%$ over NGCF, GCCF, HMLET, and LightGCN, respectively. It is worth noting that in our experimental setup, we standardized the embedding dimension for all GNN-based CF methods to 128 and conducted three rounds of feature propagation. As a result, HMLET, contrary to its original design, does not outperform LightGCN on most datasets. Moreover, we can observe that most GNN-based CF models trail slightly behind VAE-based models in performance. This indicates that GNN-based CF models might be prone to over-fitting when dealing with extensive sparse interactions, a phenomenon linked to the two issues highlighted in this paper. Besides, although DGCF employs a formally similar correlation constraint, there are significant differences between DGCF and our proposed AFDGCF, both in terms of methods and results. Methodologically, DGCF combines de-correlation with intent-aware routing to learn disentangled embeddings, aiming to ensure relative independence among learned intents. However, within each intent, certain columns remain highly correlated. From an effectiveness perspective, the DGCF method suffers from over-fitting issues, leading to noticeable performance degradation on larger datasets. Meanwhile, as mentioned in their  paper, finer-grained disentangling would lead to a performance decline in DGCF models. In these respects, our AFDGCF demonstrates a starkly different performance.

Recent studies have indicated that GNN-based CF models, like LightGCN, are much more difficult to train than traditional MF models ~\cite{peng2022svd,wang2022towards}. These models often require a larger number of epochs to reach the optimal state. We hypothesize that issues of over-correlation and over-smoothing diminish the distinction between positive and negative samples, thereby impacting learning efficiency. Fortunately, we found our AFDGCF framework can alleviate this phenomenon. As evidenced in Table~\ref{table:effi} (where "ep" means "epoch"), within the AFDGCF framework, both LightGCN and HMLET attain optimal performance with fewer epochs on various datasets. In some cases, the required epochs are reduced by half.

\vspace{-1mm}

\begin{figure*}[!t]
\centering\includegraphics[width=0.85\textwidth]{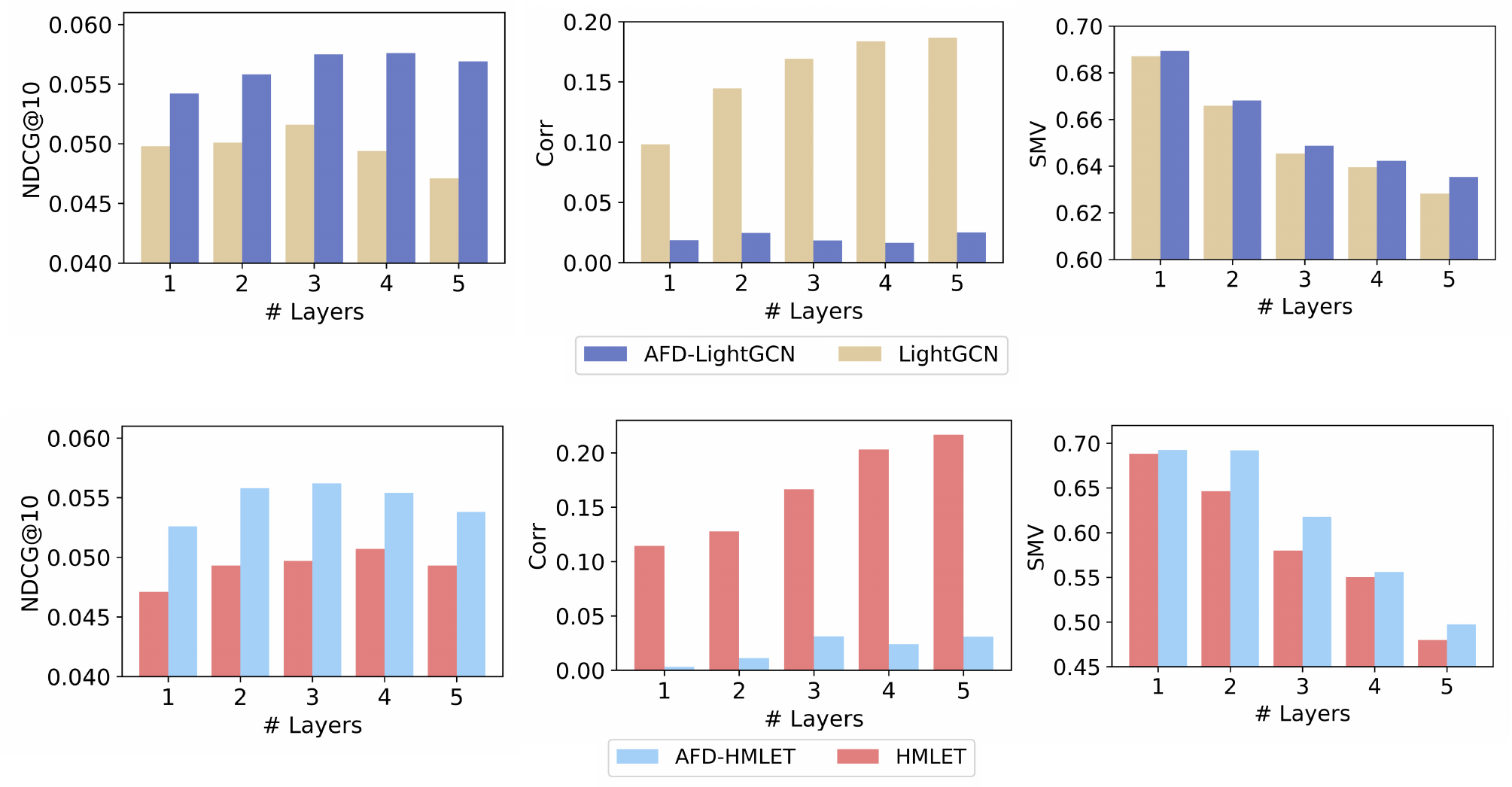}
\vspace{-4mm}
\caption{Comparison of GNN-based CF models with different layers before and after applying the AFDGCF framework in terms of NDCG@10, $\textit{Corr}$ and $\textit{SMV}$ on Yelp Dataset.}
\label{fig:case}
\vspace{-4mm}
\end{figure*}

\subsection{Ablation study of AFDGCF (RQ2)}
A key component of our proposed AFDGCF is the adaptive strategy. To validate this strategy, we conducted an ablation study to analyze its influence on performance. The experimental results are presented in Table~\ref{table:abla}, where AFD-f denotes applying a fixed penalty coefficient of $1/l$ to each layer within the AFDGCF framework. Clearly, we can observe that our adaptive approach consistently outperforms the fixed penalty coefficient across all GNN-based models, demonstrating the effectiveness of applying varying de-correlation penalties to distinct layers. Besides, we have conducted the paired \textit{t}-test to verify that the difference between each AFD and AFD-f model is statistically significant for $p < 0.05$.

\vspace{-1mm}

\subsection{Hyper-parameter Analysis (RQ3)}
In our AFDGCF framework, hyper-parameter $\alpha$ balances the CF task loss with the proposed adaptive feature de-correlation loss, acting as a regulator for the intensity of de-correlation. Figure~\ref{fig:hpyer} reports the performance of AFDGCF as $\alpha$ varies across the range $\{1e^{-5}, 5e^{-5}, \cdots, 5e^{-3}, 1e^{-2}\}$. We can observe that setting an appropriate $\alpha$ can markedly enhance AFDGCF's performance. Notably, the ideal $\alpha$ varies between datasets, but remains fairly consistent for different models within the same dataset. This is attributed to our design of applying de-correlation penalties to each layer rather than the final representation. Furthermore, for the two datasets mentioned above, we have not shown results for larger values of $\alpha$ (\textit{e.g.}, $5e^{-2}$ or $1e^{-1}$), although some model-dataset combinations may trend toward better performance at higher $\alpha$ values. However, in practice, when we apply larger values of $\alpha$, the training process of the model collapses in most cases, yielding unusable results.

\vspace{-1mm}

\subsection{Case Study (RQ4)}
We investigate the effects of our proposed AFDGCF framework using LightGCN and HMLET as examples in the following two aspects: i) the de-correlation effect on GNN-based CF models with different layers, and ii) the resulting performance enhancements.

In Figure~\ref{fig:case}, we have reported a comparison of the performance and the feature correlations of LightGCN and HMLET (with layers ranging from 1 to 5) before and after applying the AFD penalty. Observations can be made as follows: 
\begin{itemize}
    \item For any number of layers, the AFDGCF framework improves performance, and larger layers correspond to higher performance gains. These results indicate that with the increase in the number of message-passing iterations, the performance degradation of GNN-based CF models due to over-correlation and over-smoothing issues becomes more severe. However, the proposed AFDGCF framework can effectively mitigate this problem.
    \item The \textit{Corr} metric of the representations learned by AFD-LightGCN and AFD-HMLET from layers 1 to 5 is much lower than that of LightGCN and HMLET, indicating that the AFDGCF framework effectively achieves its goal of controlling the correlation between feature dimensions in the representation matrix. 
    \item The \textit{SMV} metric of the representations learned by AFD-LightGCN and AFD-HMLET from layers 1 to 5 is higher than that of LightGCN and HMLET. These results indicate that our proposed method can alleviate the over-correlation issue by constraining the column-wise correlation of the representation matrix, while simultaneously reducing the row-wise correlation of the representation matrix, thereby mitigating the over-smoothing problem.
\end{itemize} 

\vspace{-2mm}
\section{Conclusion}
This research undertook an in-depth exploration of the largely neglected issue of over-correlation in GNN-based CF models, substantiating its widespread presence and associated degradation in model performance. Our investigations established a direct and positive correlation between over-correlation and over-smoothing, highlighting their combined detrimental impact on the effectiveness of GNN-based CF models.
Specifically, we introduced the AFDGCF framework, a novel approach designed to mitigate the influence of over-correlation by strategically managing feature correlations. This was achieved through a layer-wise adaptive strategy, ensuring an optimal balance between minimizing over-smoothing and maintaining essential representation smoothness, a critical aspect for the success of GNN-based CF models.
Empirical validations underscored the efficacy of the AFDGCF framework, demonstrating notable performance enhancements across different layers of various GNN-based CF models.
\begin{acks}
This work was partially supported by National Natural Science Foundation of China (Grant No.92370204), Guangzhou-HKUST(GZ) Joint Funding Program (Grant No.2023A03J0008), Education Bureau of Guangzhou Municipality, Guangdong Science and Technology Department and Project funded by China Postdoctoral Science Foundation (Grant No.2023M730785).
\end{acks}
\bibliographystyle{ACM-Reference-Format}
\balance
\bibliography{sample-base}
\end{document}